# First principles studies of the Born effective charges and electronic dielectric tensors for the relaxor PMN (PbMg$_{1/3}$Nb$_{2/3}$O$_3$)

Narayani Choudhury[1], R. E. Cohen[2] and Eric J. Walter[3]

[1] Solid State Physics Division, Bhabha Atomic Research Centre,
Trombay, Mumbai 400 085, India

[2] Geophysical Laboratory, Carnegie Institution of Washington
5251 Broad Branch Road NW, Washington, DC 20015

[3] Department of Physics, College of William and Mary, Williamsburg, VA 23187

**Abstract**

We report first principles density functional calculations of the Born effective charges and electronic dielectric tensors for the relaxor PMN (PbMg$_{1/3}$Nb$_{2/3}$O$_3$). Visualization of the Born charge tensors as "charge ellipsoids" have provided microscopic insights on the factors governing piezoelectric enhancements with polarization rotation. Several 15 and 30-atom ferroelectric and antiferroelectric supercells of PMN involving 1:2 and 1:1 chemical ordering have been studied. A cascading set of ferroelectric phonon instabilities lead to several low symmetry monoclinic structures. We find a ground state with a 15-atom unit cell with 1:2 chemical ordering along [111] with a monoclinic *C2* structure.

## 1. Introduction

PMN(PbMg$_{1/3}$Nb$_{2/3}$O$_3$) is the relaxor end member of the solid solution PMN-PT(PbMg$_{1/3}$Nb$_{2/3}$O$_3$-PbTiO$_3$) which displays large electromechanical coupling [1] and finds application as transducers in medical ultrasound, etc. The other end member PT (PbTiO$_3$) is ferroelectric. PMN exhibits frequency dependent dielectric dispersion characteristic of relaxors and has a large, broad peak at around room temperature [1]. Relaxors are disordered over various length scales, and a multiscale approach involving a combination of various techniques including quantum mechanical, atomistic and continuum approaches [2-4] is required. The disorder in PMN gives rise to anomalies in nearly all observed properties including diffuse scattering, dynamical properties and the important applications coupled with their rich physics have resulted in extensive theoretical and experimental studies.

We have studied various ordered ferroelectric and antiferroelectric supercells for PMN. These can be used to derive interatomic potentials [2-4] and effective Hamiltonians which in turn can be employed to study the disordered solid solution PMN-PT over larger length and time scales. Our calculations reveal that ordered PMN supercells are ferroelectric.

From a materials design perspective, it is of interest to apply first principles techniques to compute the dielectric and piezoelectric properties of ferroelectric materials. The goals are to provide a microscopic understanding of the factors that can lead to enhanced performance. The key quantities that determine the piezoelectric constants are (i) the atomic response to macroscopic strain and (ii) the Born effective charge tensors. Studies of the variations of these quantities with polarization rotation are particularly of interest to understand the large piezoelectric effects [5].

We earlier reported [6] linear response studies of the long wavelength phonon modes, LO-TO (longitudinal optic-transverse optic) splittings and principal values of Born effective charges of PMN. Atomic displacements consistent with the soft mode eigenvectors obtained from a cascading set of ferroelectric zone center phonon instabilities yields various polarization rotated structures [6]. We now report the complete dynamical Born effective charges and electronic dielectric tensors for these various supercells of PMN.

## 2. Techniques

We have performed density functional LDA calculations using plane wave basis sets as implemented in the code ABINIT [7] employing well tested norm conserving pseudopotentials [6] generated using the code OPIUM [8-9]. Various ordered supercells involving 1:2 and 1:1 chemical ordering along [111] and [001] have been studied. Density functional perturbation theory linear response [10] and Berry's phase computations [11,12] were used to derive the spontaneous polarization, Born effective charge tensors, electronic dielectric tensors and phonon frequencies. The self-consistent calculations were performed with a 4x4x4 **k**-point mesh and a kinetic energy cut off of 100 Ry.

## 3. Results

### (A) Structure and energetics

PMN has a disordered perovskite based structure with an average cubic structure on a mesoscale. Although pure PMN is a relaxor, theory predicts ordered supercells to be ferroelectric [6,13-14]. We studied several 1:2 ordered 15-atom supercells ordered along [111] and [001] and found a low symmetry, ferroelectric, monoclinic *C2* ground state [6]. Prosandeev *et al.*[14] studied several complex 30-atom and 15-atom supercells, and found a 30-atom ordered, triclinic, ferroelectric, structure they denoted as [111]NT (Fig.1) to be the most stable. We also relaxed the internal atomic coordinates for this structure, but found a lower energy for a 15-atom cell with 1:2 chemical ordering and *C2* symmetry (Fig. 1) (Table I). BMN ($BaMg_{1/3}Nb_{2/3}O_3$) and SMN ($SrMg_{1/3}Nb_{2/3}O_3$) are known to be ordered 1:2 along [111]( Ref. 15) but this has never been found experimentally for PMN.

The ideal positions (as in cubic perovskite) for 1:2 chemical ordering along [111] and [001] respectively give hexagonal $P\bar{3}m1$ and tetragonal $P4/mmm$ space groups. Relaxation of the ideal hexagonal $P\bar{3}m1$ structure results in an antiferroelectric (AFE) structure [6] with Pb, Mg and three oxygen atoms having Wyckoff positions *1a*, *1b* and *3e*, respectively, having no displacements while the remaining atoms reveal antiferroelectric (AFE) displacements) (movie 1). Displacements consistent with the soft mode eigenvector of the $P\bar{3}m1$ structure, yields a ferroelectric *P3m1* structure[6] (movie 2). The hexagonal [111] FE *P3m1* structure (with spontaneous polarization along the hexagonal *c*-axis) has a doubly degenerate *E* symmetry unstable phonon mode at the zone center. Movie 3, shows the atomic displacements of the E symmetry unstable phonon eigenvector. Displacements consistent with the linear combination of this unstable eigenvector yield triclinic or monoclinic *Cm* structures. Constrained symmetry preserving relaxation gives a monoclinic *Cm (R)* structure [6] having a lower energy than the *Cm (E)* structure, obtained from the unstable mode eigenvector. The structures denoted as *Cm (R)* and *Cm (E)* monoclinic ferroelectric phases in Ref. [6] correspond to the $M_A$ and $M_B$ monoclinic structures, respectively predicted from higher order Devonshire theory [16]. All these structures are ferroelectric with larger polarization in the lower symmetry structures (Table I) and their spontaneous polarization makes an angle 0, 27 and 79 degrees with the hexagonal *c*-axis for the hexagonal *P3m1*, and monoclinic *Cm ($M_B$)* and *Cm ($M_A$)* structures, respectively. The *Cm ($M_A$)* structure has a single unstable long wavelength phonon mode and the ground state is found to have a monoclinic *C2* structure. The rotation of polarization in going from ideal $P\bar{3}m1$ → *P3m1* → *Cm ($M_B$)* → *Cm ($M_A$)* → *C2* → ideal $P\bar{3}m1$ is evident from movie 4.

### (B). Born effective charges, electronic dielectric tensors and phonon frequencies

The computed Born effective charge tensors for the [111] ordered supercells are given in Tables II and III. Although the principal values of the Born tensors for these structures (except the monoclinic *Cm($M_B$)* and *C2* structures) have been reported [6], the information about the principal axes directions necessary to derive the complete tensors have not been reported. The anisotropy of the charge tensors has an important bearing on their physical properties and the complete tensors and principal values are useful for microscopic interpretation of piezoelectric data. The monoclinic FE structures studied in Table II, also involve polarization rotation and their relative variations are of additional interest.

The Born tensors are not completely symmetric in the [111] ordered supercells, but symmetrization of the charge tensors does not change their polar properties such as the LO-TO splittings significantly [6]. Thus we symmetrized the effective charge tensors and visualized them as "charge ellipsoids". The symmetrized dynamic Born effective charge tensors for the *P3m1* FE and the AFE *P$\bar{3}$m1* structures are displayed in Fig. 2. The effective charges of all the atoms in the [111] ordered structures are significantly larger than their nominal ionic values, as typically obtained for other ferroelectric perovskites [18], due to the covalent character of the B-O bonds and the large polarizability of the oxygen ions. The computed charge tensors in the AFE *P$\bar{3}$m1* structure (Table II) are in good agreement with the reported values for the [111]NNM supercells of Prosandeev *et al.* [14].

The charge tensors in the [111] AFE *P$\bar{3}$m1* and [111]FE *P3m1* structures are quite similar. The magnitude of the Born effective charges of the oxygen atoms are significantly larger along the Nb-O bonds as compared to other directions, due to the covalency of the Nb-O bonds. The anisotropy of the charge tensors have important consequences and play an important role in the piezoelectric enhancements which occur in directions which maximize the charge contributions from various atoms along with ferroelectric displacements. These can explain the enhancements involving polarization rotation and are discussed in detail in Ref [6]. The Born effective charge tensors for the low symmetry [111] *Cm ($M_B$)*, *Cm ($M_A$)*, and *C2* structure are displayed in Fig. 3. Although, the charge values in these different structures are similar, the charge ellipsoids have different orientations. These have an important bearing on their their polarization and piezoelectric response. The animation of the changes in orientations of the Born tensors involving these various ferroelectric supercells involving polarization rotation is displayed in Movie 5.

The Born effective charge tensors in the [001] ordered tetragonal *P4/mmm* structure is completely symmetric and the vanishing non-diagonal elements for all the atoms result in the ellipsoids having principal axes, which are along the tetragonal **a**, **b**, and **c** directions (Fig. 2). The Born charge tensors [6] for the tetragonal *P4/mmm* structure have average value of 3.73, 2.14, 8.16, -2.2, -5.4 for the Pb, Mg, Nb and O $\perp$ and O$_{\parallel}$, respectively. The charge tensors and principal values particularly, for the Mg, Nb and O are significantly different for the [111] and [001] ordered supercells (Figs2-3). These lead to differences in their vibrational spectra (Fig. 4) and polar response [6].

The computed electronic dielectric tensors for the [111] ordered supercells are given in Table IV. Significant variations are seen for the [111] ordered structures involving polarization rotation. The computed electronic dielectric constants using the LDA approximation are overestimated as compared with the experimental results [13-14] of about 5.83 for all these supercells.

The computed long wavelength phonon frequencies for these various 1:2 chemically ordered supercells are compared with observed Raman [14,19] and infrared data [14] in Fig. 4. All these supercells ordered along [111] except the *C2* have a ferroelectric instability around 45i which is stabilized in the *C2* structure. The calculated phonon frequencies are in overall good agreement with experiments[14,19], especially as these are idealized supercells. The ground state *C2* phonon frequencies have been evaluated both at the experimental volume (V=1343 Bohr$^3$) and at a constant volume V=1304 Bohr$^3$ used to study the various other supercells [6]. The volume variations of the phonon frequencies for the *C2* structure for these volumes are small. The phonon instabilities in PMN have an important bearing on their dielectric and piezoelectric properties. The computed phonon spectra of the ground state *C2* structure are overall similar to the spectra obtained for the various low symmetry [111] ordered structures. Several polar phonon modes show large LO-TO splittings and have been analyzed in detail in Ref. 6. These results will be useful to derive interatomic potentials required for studying relaxors on larger length scales.

*Acknowledgements*
This work was supported by the Office of Naval Research under contract number N000149710052. We thank Z.Wu for helpful discussions. Computations were performed at the Center for Piezoelectrics by Design (CPD), College of William and Mary.

**TABLE I**: The space group, relative energies (mRy/15-atom cell), and lattice constants (Bohr) for various ferroelectric 1:2 ordered supercells of PMN ordered along [111] (V=1304 Bohr$^3$). The ground state has a monoclinic *C2* structure. The [111]NT is a complex 1:1 ordered 30 atom triclinic supercell, obtained by relaxing the reported [111]NT ground state structure of Prosandeev *et al.* [13-14]. The monoclinic *Cm* ($M_B$) structure was obtained from the *E* unstable phonon mode eigenvector in the ferroelectric *P3m1* structure, while the *Cm* ($M_A$) structure was obtained via constrained structural relaxation. $P_x$, $P_y$, $P_z$ are the polarization (C/m$^2$) along Cartesian directions. The polarization *P* has been estimated from the atomic displacements with respect to ideal positions and average values of Born effective charges (using Z(Pb)=3.74, Z(Mg)=2.85, Z(Nb)=6.76, Z(O)=-3.2). For the hexagonal [111]FE *P3m1* structure, Berry's phase calculations using 4x4x20 k-point mesh gives a value 0.52 C/m$^2$ in good agreement with these estimations. The angle (degrees) is specified with respect to direction of spontaneous polarization of the hexagonal *P3m1* structure along [-1-1-1].

| Space group | ΔE mRy | a | b | c | α | β | γ | $P_x$ | $P_y$ | $P_z$ | \|P\| | Angle |
|---|---|---|---|---|---|---|---|---|---|---|---|---|
| *P3m1* | 0 | 10.7127 | 10.7127 | 13.1203 | 90 | 90 | 120 | -.32 | -.32 | -.32 | 0.55 | 0 |
| *Cm* ($M_B$) | -1.11 | 10.7127 | 10.7127 | 13.1203 | 90 | 90 | 120 | -.44 | -.44 | -.09 | 0.63 | 27.26 |
| *Cm* ($M_A$) | -7.47 | 10.635 | 10.769 | 13.096 | 89.78 | 89.97 | 119.62 | -.31 | -.31 | 0.52 | 0.67 | 85.90 |
| *C2* | -9.36 | 10.677 | 10.830 | 13.085 | 90.03 | 89.43 | 120.48 | -.50 | -.00 | 0.50 | 0.70 | 90 |
| [111]NT *P1* | -5.59 | 18.494 | 18.506 | 10.653 | 73.22 | 73.18 | 119.30 | -.32 | -.30 | 0.44 | 0.62 | 80.82 |

**Table II:** The dynamic Born effective charge tensors (Cartesian coordinates) for the [111] ordered supercells in PMN. $x$, $y$ and $z$ are along pseudocubic perovskite directions. The Wyckoff positions (WP) are shown and only the charges for the inequivalent atoms are listed. The $P\bar{3}m1$ structure is antiferroelectric, while the $P3m1$, $Cm(M_B)$, $Cm(M_A)$ and $C2$ supercells are ferroelectric. The unit cell constants are given in Table I. The complete structures are given in Ref. [6].

| Space group | Atom Type | WP | $Z^*_{xx}$ | $Z^*_{yy}$ | $Z^*_{zz}$ | $Z^*_{xy}$ | $Z^*_{xz}$ | $Z^*_{yx}$ | $Z^*_{yz}$ | $Z^*_{zx}$ | $Z^*_{zy}$ |
|---|---|---|---|---|---|---|---|---|---|---|---|
| $P3m1$ | Pb | 1a | 3.79 | 3.79 | 3.79 | -.37 | -.37 | -.37 | -.37 | -.37 | -.37 |
| | Pb | 1b | 3.57 | 3.57 | 3.57 | .48 | .48 | .48 | .48 | .48 | .48 |
| | Pb | 1c | 3.62 | 3.62 | 3.62 | -.18 | -.18 | -.18 | -.18 | -.18 | -.18 |
| | Mg | 1a | 2.85 | 2.85 | 2.85 | -.01 | -.01 | -.01 | -.01 | -.01 | -.01 |
| | Nb | 1b | 6.72 | 6.72 | 6.72 | -.15 | -.15 | -.15 | -.15 | -.15 | -.15 |
| | Nb | 1c | 6.53 | 6.53 | 6.53 | -.31 | -.31 | -.31 | -.31 | -.31 | -.31 |
| | O | 3d | -3.59 | -2.48 | -2.48 | -.12 | -.12 | -.01 | -.09 | -.01 | -.09 |
| | O | 3d | -2.30 | -4.01 | -2.30 | .46 | -.24 | .20 | .20 | -.24 | .46 |
| | O | 3d | -2.21 | -2.21 | -5.50 | -.03 | .07 | -.03 | .07 | .31 | .31 |
| $P\bar{3}m1$ | Pb | 1a | 3.91 | 3.91 | 3.91 | -.43 | -.43 | -.43 | -.43 | -.43 | -.43 |
| | Pb | 2d | 3.97 | 3.97 | 3.97 | .21 | .21 | .21 | .21 | .21 | .21 |
| | Mg | 1b | 2.85 | 2.85 | 2.85 | .03 | .03 | .03 | .03 | .03 | .03 |
| | Nb | 2d | 6.74 | 6.74 | 6.74 | -.21 | -.21 | -.21 | -.21 | -.21 | -.21 |
| | O | 6i | -3.91 | -2.51 | -2.51 | -.02 | -.02 | .15 | -.02 | .15 | -.02 |
| | O | 3e | -2.37 | -2.37 | -5.59 | .05 | -.11 | .05 | -.11 | .25 | .25 |
| $Cm(M_A)$ | Pb | 1a | 3.69 | 3.69 | 3.30 | -.42 | -.39 | -.42 | -.39 | -.22 | -.22 |
| | Pb | 1a | 3.84 | 3.84 | 3.64 | -.05 | .40 | -.05 | .40 | .50 | .50 |
| | Pb | 1a | 3.50 | 3.50 | 3.62 | .45 | .08 | .45 | .08 | -.19 | -.19 |
| | Mg | 1a | 2.87 | 2.87 | 2.90 | .01 | .07 | .01 | .07 | .08 | .08 |
| | Nb | 1a | 6.53 | 6.53 | 5.87 | -.09 | -.26 | -.09 | -.26 | .02 | .02 |
| | Nb | 1a | 6.12 | 6.12 | 6.49 | -.40 | .11 | -.40 | .11 | -.14 | -.14 |
| | O | 2c | -3.50 | -2.58 | -2.40 | -.18 | .22 | -.03 | .19 | .34 | .09 |
| | O | 1a | -2.14 | -2.14 | -4.05 | -.18 | -.22 | -.18 | -.22 | -.28 | -.28 |
| | O | 2c | -2.31 | -3.98 | -2.17 | .44 | .14 | .25 | -.34 | .13 | -.29 |
| | O | 1a | -2.47 | -2.47 | -3.50 | -.11 | .27 | -.11 | .27 | .16 | .16 |
| | O | 1a | -2.15 | -2.15 | -4.97 | -.13 | -.41 | -.13 | -.41 | -.02 | -.02 |
| | O | 2c | -2.24 | -5.19 | -2.08 | .09 | .15 | .35 | -.01 | .22 | -.40 |
| $Cm(M_B)$ | Pb | 1a | 3.66 | 3.66 | 3.85 | -.39 | -.38 | -.39 | -.38 | -.27 | -.27 |
| | Pb | 1a | 3.48 | 3.48 | 3.77 | .40 | .49 | .40 | .49 | .53 | .53 |
| | Pb | 1a | 3.37 | 3.37 | 3.89 | .04 | -.14 | .04 | -.14 | -.36 | -.36 |
| | Mg | 1a | 2.86 | 2.86 | 2.83 | -.08 | .02 | -.08 | .02 | .02 | .02 |
| | Nb | 1a | 6.68 | 6.68 | 6.16 | -.01 | -.15 | -.01 | -.15 | -.10 | -.10 |
| | Nb | 1a | 6.29 | 6.29 | 6.62 | -.18 | -.29 | -.18 | -.29 | -.32 | -.32 |
| | O | 2c | -3.44 | -2.42 | -2.63 | -.29 | -.01 | -.12 | -.03 | .09 | -.09 |
| | O | 1a | -2.28 | -2.28 | -3.71 | -.12 | -.02 | -.12 | -.02 | -.12 | -.12 |
| | O | 2c | -2.15 | -4.00 | -2.28 | .48 | -.19 | .19 | .15 | -.17 | .35 |
| | O | 1a | -2.39 | -2.39 | -3.93 | -.29 | .46 | -.29 | .46 | .26 | .26 |
| | O | 1a | -2.22 | -2.22 | -5.14 | -.22 | -.15 | -.22 | -.15 | .13 | .13 |
| | O | 2c | -2.00 | -5.42 | -2.26 | .23 | .05 | .37 | .19 | .13 | -.10 |
| $C2$ | Pb | 1a | 3.34 | 3.93 | 3.35 | -0.29 | -0.24 | -0.50 | -0.50 | -0.23 | -0.28 |
| | Pb | 2c | 3.57 | 3.79 | 3.50 | -0.13 | 0.00 | 0.09 | 0.58 | 0.24 | 0.53 |
| | Mg | 1b | 2.89 | 2.87 | 2.88 | 0.04 | 0.14 | 0.02 | 0.02 | 0.14 | 0.04 |
| | Nb | 2c | 6.39 | 6.41 | 5.70 | 0.03 | -0.39 | 0.11 | -0.26 | -0.13 | -0.08 |
| | O1 | 2c | -3.39 | -2.74 | -2.32 | -0.11 | 0.43 | 0.05 | 0.12 | 0.43 | 0.03 |
| | O2 | 2c | -1.95 | -2.19 | -4.03 | -0.06 | -0.19 | -0.04 | 0.03 | -0.19 | -0.10 |
| | O3 | 2c | -2.39 | -3.69 | -2.34 | -0.20 | 0.16 | -0.35 | 0.19 | 0.15 | 0.33 |
| | O4 | 2c | -2.08 | -2.27 | -4.79 | -0.03 | -0.56 | -0.10 | -0.19 | -0.14 | 0.12 |
| | O5 | 1a | -2.09 | -5.37 | -2.07 | -0.13 | 0.31 | 0.20 | 0.19 | 0.31 | -0.15 |

**Table III:** The principal values of dynamic Born tensors for the monoclinic $Cm$ ($M_B$) and $C2$ structures.

|          | Principal Value 1 | Principal Value 2 | Principal Value 3 |
|----------|-------------------|-------------------|-------------------|
| $Cm$ ($M_B$) |               |                   |                   |
| Pb       | 3.01              | 4.05              | 4.10              |
| Pb       | 3.08              | 3.10              | 4.55              |
| Pb       | 3.22              | 3.33              | 4.07              |
| Mg       | 2.77              | 2.85              | 2.93              |
| Nb       | 6.11              | 6.70              | 6.72              |
| Nb       | 5.87              | 6.47              | 6.86              |
| O1       | -3.48             | -2.64             | -2.37             |
| O2       | -3.71             | -2.40             | -2.16             |
| O4       | -4.10             | -2.32             | -2.02             |
| O5       | -4.11             | -2.50             | -2.10             |
| O7       | -5.14             | -2.45             | -2.00             |
| O8       | -5.45             | -2.29             | -1.95             |
|          |                   |                   |                   |
| $C2$     |                   |                   |                   |
| Pb       | 2.83              | 3.58              | 4.21              |
| Pb       | 3.05              | 3.59              | 4.23              |
| Mg       | 2.74              | 2.85              | 3.04              |
| Nb       | 5.58              | 6.33              | 6.57              |
| O1       | -3.54             | -2.75             | -2.16             |
| O2       | -4.05             | -2.19             | -1.93             |
| O3       | -3.81             | -2.41             | -2.21             |
| O4       | -4.84             | -2.29             | -2.02             |
| O5       | -5.38             | -2.39             | -1.77             |

**Table IV:** The computed electronic dielectric tensors (Cartesian coordinates) for the various 1:2 ordered supercells along [111] in PMN. x, y and z are along pseudocubic directions. The dielectric tensors are symmetric with $\varepsilon_{ij}=\varepsilon_{ji}$.

|                | $\varepsilon_{xx}$ | $\varepsilon_{yy}$ | $\varepsilon_{zz}$ | $\varepsilon_{xy}$ | $\varepsilon_{xz}$ | $\varepsilon_{yz}$ |
|----------------|--------------------|--------------------|--------------------|--------------------|--------------------|--------------------|
| $P3m1$         | 6.67               | 6.67               | 6.67               | -0.12              | -0.12              | -0.12              |
| $P\bar{3}m1$   | 7.13               | 7.13               | 7.13               | -0.11              | -0.11              | -0.11              |
| $Cm$ ($M_A$)   | 6.52               | 6.52               | 6.42               | -0.11              | 0                  | 0                  |
| $Cm$ ($M_B$)   | 6.49               | 6.49               | 6.61               | -0.06              | -0.10              | -0.10              |
| $C2$           | 6.31               | 6.56               | 6.31               | 0.003              | -0.071             | 0.003              |

**Fig. 1 (Color online):** Polyhedral representation of the supercells in PMN. (Top) We find a lowest energy for the monoclinic *C2* 15-atom supercell with 1:2 chemical ordering along [111]. (Bottom) The triclinic 30-atom [111]NT supercell found by Prosendeev *et al* [14] as the ground state.

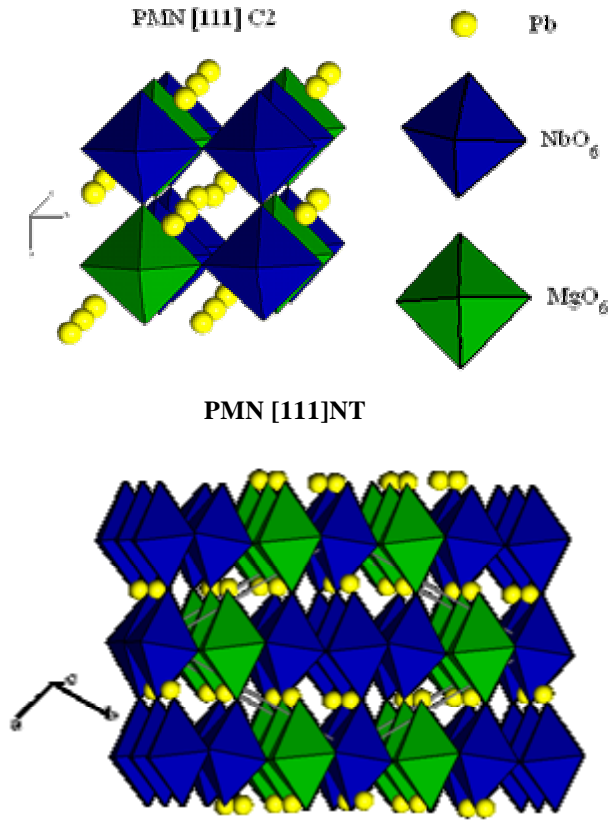

**Fig. 2 (Color online):** Quadric representation of the Born effective charge tensors as "charge ellipsoids" for the various supercells with 1:2 chemical ordering along [111] and [001], generated using the software xtaldraw [16]. The [111] FE (space group $P3m1$) and [111] AFE (space group $P\bar{3}m1$) are hexagonal. (a), (c) give the view with the hexagonal **c**-axis as vertical, while (b) and (d) are in the *ab*-plane, viewed down **c**. The [001] ordered antiferroelectric supercell is tetragonal (space group $P4/mmm$) and (e) gives the view with the tetragonal **c**-axis as vertical while (f) gives the view in the *ab*-plane looking down the **c**-axis. The charge tensors for the oxygen atoms in the hexagonal [111]FE and [111]AFE structures were symmetrized for the display.

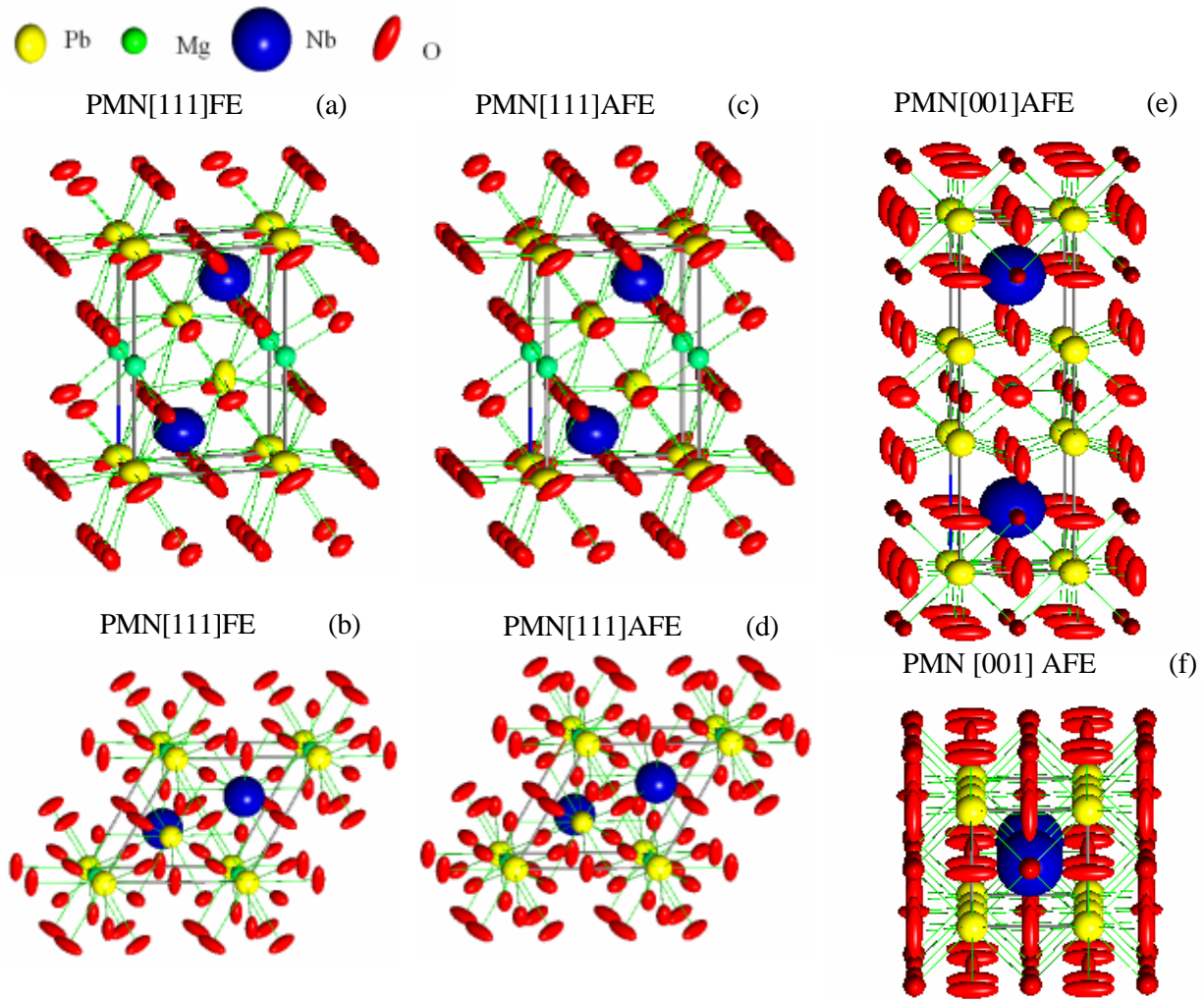

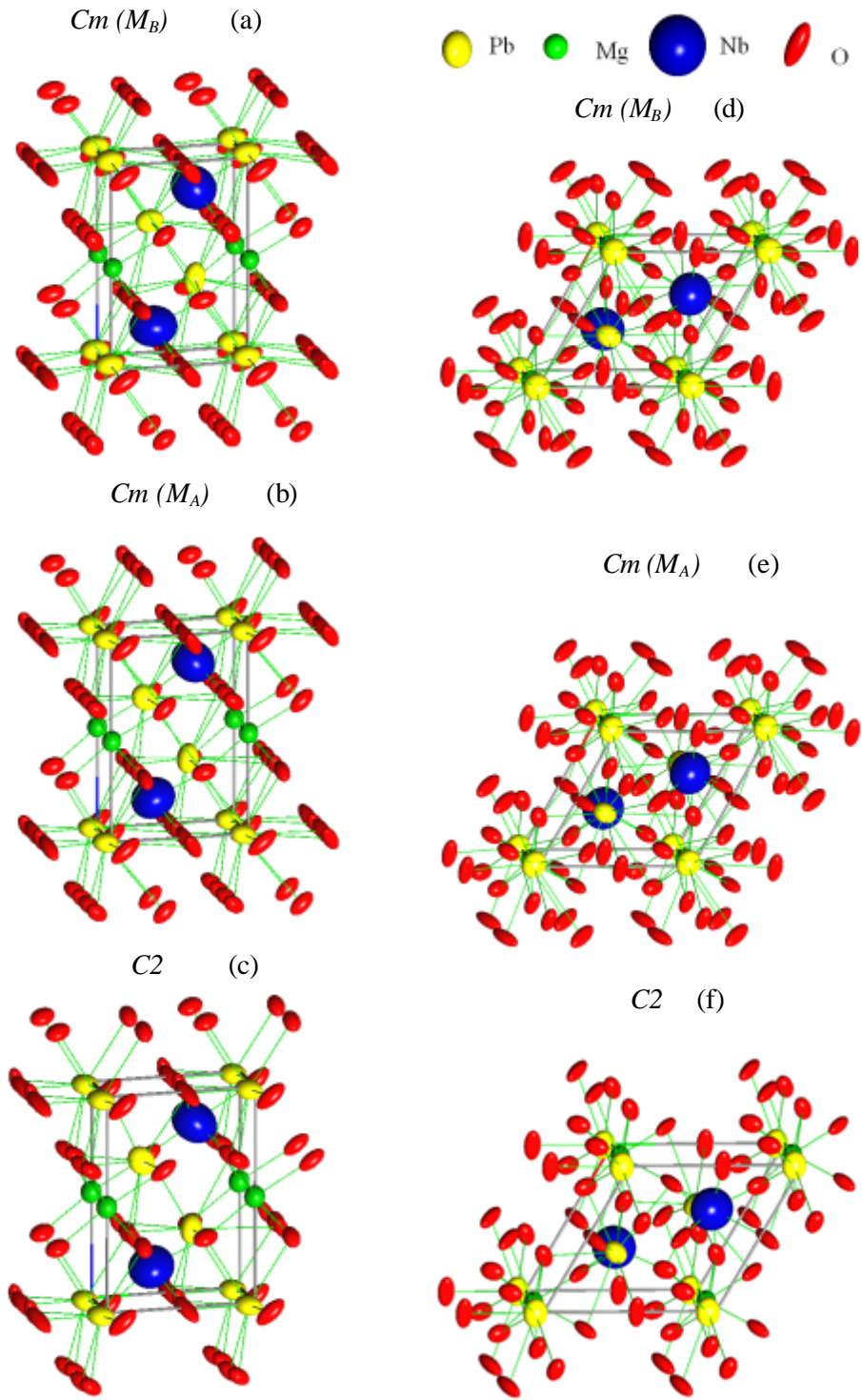

**Fig. 3 (Color online):** Quadric representation of the symmetrized Born effective charge tensors as "charge ellipsoids" for the various monoclinic supercells with 1:2 chemical ordering along [111], generated using the software xtaldraw [17]. (a), (b) and (c) give the view with $c$-axis vertical, while (e), (f) and (g) give the view in the basal plane, viewed down the $c$-axis.

**Fig. 4**. The computed phonon frequencies for the various 1:2 chemically ordered supercells along [111] compared with available experimental Raman[14,19] and infrared [14] data. Unstable phonon frequencies are depicted as negative frequencies. All the long wavelength phonon modes in the ground state *C2* structure are stable.

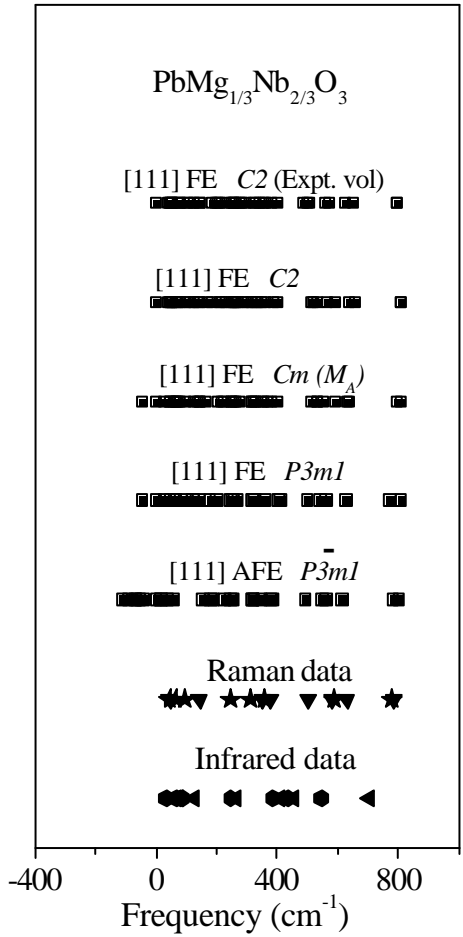

**Movie 1.** Animation of the AFE displacements in the $P\bar{3}m1$ structure (hexagonal **c**-axis is vertical, and **a**-axis horizontal). The Pb (yellow), Mg (blue), Nb (green) and O (red) atoms are shown.

**Movie 2.** The atomic displacements involving the AFE $P\bar{3}m1$ to FE *P3m1* lattice instability (hexagonal **c**-axis is vertical, and **a**-axis horizontal). The various colors represent Pb (yellow), Mg (blue), Nb (green) and O (red).

**Movie 3.** The eigenvector of the doubly degenerate E symmetry 45i unstable TO phonon mode in the FE *P3m1* structure (hexagonal **c**-axis is vertical, and **a**-axis horizontal). While the atomic displacements shown are along the hexagonal **a**-axis, the degenerate mode has identical displacements along the **b**-axis. The monoclinic *Cm* ($M_B$) structure is obtained by displacing the atoms using a linear combination of these eigenvectors. The various colors represent Pb (yellow), Mg (blue), Nb (green) and O (red).

**Movie 4.** Animation of the atomic displacements with polarization rotation (hexagonal **c**-axis is vertical, and **a**-axis horizontal). A closed loop display of the displacements in going from ideal → *P3m1* → *Cm* ($M_B$) → *Cm* ($M_A$) → *C2* → ideal is shown. The various colors represent Pb (yellow), Mg (green), Nb (blue) and O (red).

**Movie 5.** Animation of the changes in Born charge tensor ellipsoids involving polarization rotation (hexagonal **c**-axis is vertical, and **a**-axis horizontal). A closed loop display of the changes in going from *P3m1* → *Cm* ($M_B$) → *Cm* ($M_A$) → *C2* → *P3m1* is shown. The various colors represent Pb (yellow), Mg (green), Nb (blue) and O (red).